\documentclass[aps,twocolumn,showpacs,preprintnumbers,amsmath,amssymb,floatfix]{revtex4}
\usepackage{graphicx}
\usepackage{dcolumn}

\begin{document} 
\title{Interference between Extrinsic and Intrinsic Losses in XAFS} 

\author{L. Campbell,$^1$ L. Hedin,$^2$ J. J. Rehr,$^1$ and
W. Bardyszewski$^3$} 

\affiliation{$^1$Dept.~of Physics, Univ.~of Washington, Seattle, Washington } 

\affiliation{$^{2}$Dept.~of Physics, Lund University, Lund, Sweden, and MPI-FKF, 
Stuttgart, Germany;} 

\affiliation{$^3$Dept.~of Physics, Univ.~of Warsaw, Warsaw, Poland} 

\date{\today} 
 
\begin{abstract} 
The interference between extrinsic and intrinsic losses in x-ray absorption 
fine structure (XAFS) is treated within a Green's function formalism, 
without explicit reference to final states. The approach makes use of a 
quasi-boson representation of excitations and perturbation theory in the 
interaction potential between electrons and quasi-bosons. These losses lead 
to an asymmetric broadening of the main quasi-particle peak plus an 
energy-dependent satellite in the spectral function. The x-ray absorption 
spectra (XAS) is then given by a convolution of an effective spectral 
function over a one-electron cross-section. It is shown that extrinsic and 
intrinsic losses tend to cancel near excitation thresholds, and 
correspondingly, the strength in the main peak increases. At high energies, 
the theory crosses-over to the sudden approximation. These results thus 
explain the observed weakness of multi-electron excitations in XAS. The 
approach is applied to estimate the many-body corrections to XAFS, beyond 
the usual mean-free-path, using a phasor summation over the spectral 
function. The asymmetry of the spectral function gives rise to an additional 
many-body phase shift in the XAFS formula. 
\end{abstract} 
 

\maketitle 
 
\section{Introduction} 
 
The treatment of inelastic losses in x-ray absorption fine structure (XAFS) 
has long been of interest \cite{Rehr78,Bardyszewski87,Hedin89}. Two types of 
losses are identified. {\it Extrinsic\/} losses occur during the propagation 
of the photoelectron, and are caused by the creation of excitations such as 
plasmons, electron-hole pairs, etc. {\it Intrinsic\/} losses refer to the 
creation of excitations by the sudden appearance of the core hole. The 
intrinsic losses are often called shake-up and shake-off excitations but are 
of the same type as the extrinsic, i.e.\ plasmons, multi-electron 
excitations, etc. Typically these excitations are observed only weakly in 
x-ray absorption spectra (XAS). The extrinsic losses cause a decrease in 
intensity in the no-loss or primary channel, which is usually treated 
phenomenologically in terms of a mean-free path $\lambda $.
Owing to the 
difficulty of quantitative calculations, these additional losses have usually 
been neglected or represented as a constant amplitude factor, on the 
understanding that they only give some smooth background. 
 
The question of possible {\it interference \/} between extrinsic and 
intrinsic losses has long been unsettled. For photoemission spectroscopy, it 
has been shown \cite{Inglesfield83,Bardyszewski85} that this interference is 
particularly important near excitation thresholds, where the losses strongly 
cancel. This cancellation results from the opposite signs of the coupling 
between the photoelectron and the core-hole to excited states of the valence 
electrons. In particular, for plasmon creation at threshold, only the 
long-wavelength plasmons appear which, due to momentum conservation, makes 
the recoil of the electron in the extrinsic losses small. The intrinsic 
losses are caused by coupling to the core hole, which cannot recoil. As a 
result the extrinsic and intrinsic couplings become equal at threshold. The 
situation in XAFS is less clear. Fujikawa has discussed this cancellation 
using an explicit calculation of matrix elements and approximate closure 
relations \cite{Fujikawa93}. However, his discussion is limited to the 
vicinity of threshold, while we find that the cancellation effects extend 
over a wide energy region. 
 
In this paper we present a formal analysis of the loss problem, which is an 
extension of work by Hedin and Bardyszewski \cite 
{Bardyszewski87,Hedin89,Bardyszewski85}, together with numerical calculations 
to illustrate the theory. The results here are formulated in terms of an 
effective one-particle propagator which includes both inelastic losses and 
interference effects. This propagator contains an asymmetric quasi-particle 
peak plus a broad energy dependent satellite structure. Our approach is 
essentially a generalization of the GW approximation which, in addition to 
extrinsic losses, treats {\it intrinsic} losses and {\it interference} 
terms. The formalism also partly accounts for edge-singularity effects and 
contains corrections to the final state rule. Although the cumulant expansion
was successfully used to describe {\it intrinsic\/} losses in valence electron
photoemission \cite{Aryasetiawan96}, we have found it more 
difficult to apply for x-ray absorption spectra. For photoemission (PES) the 
propagator for a hole and the spectral function $A^{h}(\omega )=A(\omega 
)\theta (E_{F}-\omega )$ are needed, while in XAS we need the propagator for 
a particle and the spectral function $A^{p}(\omega )=A(\omega )\theta 
(\omega -E_{F})$. The GW approximation for $A(\omega )$ was discussed in
Ref.~{\onlinecite{Hedin69} }(p 87, 92). The satellite of the hole spectral 
function is very strong and sharp at the bottom of the band, but decreases 
in intensity and broadens as one approaches the Fermi energy$E_{F}$ \cite 
{Hedin69,Hedin80}. At $E_{F}$, the satellite for the particle spectrum is 
similar to that for the hole spectrum, and it rapidly becomes weaker and 
broader with increasing energy above $E_{F}$. The GW approximation places 
the satellite at about $1.5\;\omega _{p}$, rather than at $\omega _{p}$ away 
from the quasi-particle peak, as is predicted by the cumulant approximation, 
and born out by PES experiments \cite{Aryasetiawan96}. However, above the 
Fermi level the difference between the GW and cumulant approximations is 
less pronounced. 
 
Our formulation clarifies the nature of inelastic losses in XAS
and also yields semi-quantitative estimates of their effects, based 
on the electron gas approximation. In particular the theory yields an 
estimate for the reduction in the XAFS amplitude due to inelastic losses in 
terms of a phasor summation over the spectral function. This results in an 
energy dependent reduction factor $|S_0^2(\omega)|$ to the usual XAFS formula, 
as well as an additional many-body phase shift. Near the excitation 
threshold, we find that there is appreciable cancellation of strong 
extrinsic and intrinsic losses by the interference terms. Correspondingly, 
the strength of the primary channel, i.e., the main quasi-particle peak, 
increases near threshold. Thus the theory also explains the 
surprising weakness of multi-electron excitations in the observed XAS \cite 
{Rehr00}. At sufficiently high energies both the extrinsic and the 
interference contributions become negligible, and the theory crosses over to 
the sudden-approximation limit, where only intrinsic losses remain. 
Our theory is illustrated for the case of Cu metal. This system provides a
good test case for our theory since errors in the conventional MS
expansion and potentials are minimal and accurate experimental data is available
\cite{Rehr92}.

\section{Theory} 
 
\subsection{Basic expressions} 
 
The x-ray absorption coefficient $\mu(\omega)$ can be expressed formally in 
terms of the many-electron Green's function $G(E)=1/(E-H+i\gamma)$ as  
\begin{equation} 
\mu \left( \omega \right) =-{\frac{1}{\pi }}{\rm Im}\,\left\langle \Psi 
_{0}\left| \Delta ^{\dagger }\,{\frac{1}{E_{0}+\omega -H+i\gamma }}\Delta 
\right| \Psi _{0}\right\rangle ,  \label{mu1} 
\end{equation} 
where $|\Psi _{0}\rangle $ is the $N$-particle ground state of the total 
system (valence electrons and ion cores), $H$ the Hamiltonian which fully 
includes electron-electron interactions, $E_{0}$ the ground state energy (we 
consider for simplicity only the case when the temperature is zero), and $%
\omega $ the photon energy. We use atomic units $m=\left| e\right| =\hbar 
=1 $, where lengths are in Bohr radii (0.529\AA) and energies in Hartrees 
(27.2 eV). Further  
\begin{equation} 
\Delta =\Sigma _{k}\langle k|d|b\rangle c_{k}^{\dagger }b+hc  \label{delta} 
\end{equation} 
is the dipole operator coupling the photon to the electronic system, and $%
\gamma $ the inverse core hole lifetime. We have assumed that a specific 
core level $\left| b\right\rangle $ on a specific atom is involved.
 
The standard way to proceed from Eq.~(\ref{mu1}) is to insert a complete set 
of interacting states $\left| \Psi _{n}\right\rangle $, which recovers the 
golden rule expression when we take $\gamma $ as infinitesimal; a finite $%
\gamma $ just gives a Lorentzian broadening, i.e.,  
\begin{equation} 
\mu \left( \omega \right) =\sum_{n}\left| \langle \Psi _{n}|\Delta |\Psi 
_{0}\rangle \right| ^{2}\delta (E_{0}+\omega -E_{n}).  \label{goldenrule} 
\end{equation} 
This expression contains explicit final states $\left| \Psi 
_{n}\right\rangle $ in which the excited electron (the photoelectron) is 
correlated with the valence electrons. Such states are very difficult to 
handle, so we instead take a different route which leads to an expression 
where no explicit final states are involved but instead, an expansion in 
one electron Green's functions, as in conventional XAS theory. Our
formulation, however, contains shake up effects and extrinsic inelastic
losses as well as interference terms, and not just the damping in the
elastic channel.

Some of the formal results derived here were presented previously
in short conference reports \cite{Bardyszewski87,Hedin89}.  In this paper
we have concentrated on the shake-up effects of the core
hole potential and on the extrinsic losses of the ejected electron.
To have a clean picture, we have regarded the core electron as
structureless. Thus we have not considered interesting problems like
core hole degeneracies and L$_{II}$/L$_{III}$ edges. Such problems are in
principle complicated Kondo type problems. If we, however, limit the
treatment to have an ion embedded in a solid state environment, and only
consider the multiplet effects for the embedded ion \cite{Sawatzky91},
we think our treatment could straightforwardly be extended to this situation.

We take for $H$ the standard ``deep-level'' Hamiltonian discussed e.g.,
by Langreth \cite{Langreth70} (cf.\ Ref.\ \onlinecite{Almbladh83}, pp 645),  
\begin{equation} 
H=H_{v}+\varepsilon _{c}b^{\dagger }b+Vbb^{\dagger }.  \label{HLangreth} 
\end{equation} 
Here $H_{v}$ describes the electrons outside the ion cores, i.e., the 
valence electrons and the photoelectron, and $V$ the interactions between 
the outer electrons and the core hole. This approximation neglects virtual 
excitations of the core electrons, but takes correlations among the outer 
electrons and the potential $V$ from the core hole fully into account. 
Neglecting core-valence exchange we have explicitly,  
\begin{equation} 
V=-\sum_{i}w({\bf r}_{i}),\;\quad w({\bf r})=\int v({\bf r-r}^{\prime })\rho 
_{b}({\bf r}^{\prime })d{\bf r}^{\prime },  \label{Vc} 
\end{equation} 
where $v({\bf r})$ is the Coulomb potential, and $\rho _{b}({\bf r})$ the 
charge density of the core electron state ``$b$''. The ground state of $H$
$\left| \Psi _{0}\right\rangle $, is thus a product,  
\begin{equation} 
\left| \Psi _{0}\right\rangle =\left| \Phi _{0}\right\rangle \left| 
b\right\rangle ,  \label{GS1} 
\end{equation} 
where $\left| \Phi _{0}\right\rangle $ is the correlated wave function for $%
N_{v}$ outer electrons and $\left| b\right\rangle $\ is the core electron 
wave function, 
 
\[ 
H_{v}\left| \Phi _{0}\right\rangle =E_{0}^{0}\left| \Phi _{0}\right\rangle 
,\;H\left| \Psi _{0}\right\rangle =E_{0}\left| \Psi _{0}\right\rangle 
,\;E_{0}=E_{0}^{0}+\varepsilon _{c}.  
\] 
The passive core electrons are not written out explicitly. The core 
electrons are easily eliminated, and Eqs.~(\ref{mu1}), (\ref{delta}) and (%
\ref{GS1}) give  
\begin{eqnarray} 
\mu (\omega ) &=&-{\frac{1}{\pi }}\,{\rm Im}\,\sum_{k_{1}k_{2}}\langle 
b|d^{\dagger }|k_{1}\rangle \langle k_{2}|d|b\rangle  \nonumber \\ 
&\times &\langle \Phi _{0}|c_{k_{1}}\frac{1}{E_{0}+\omega -H_{v}^{\prime 
}+i\gamma }c_{k_{2}}^{\dagger }|\Phi _{0}\rangle ,  \label{mu3} 
\end{eqnarray} 
where  
\[ 
H_{v}^{\prime }=H_{v}+V.  
\] 
Here and elsewhere in this paper, we will use a prime to denote quantities 
calculated in the presence of a core hole. The x-ray absorption is now 
formally given by a one electron expression  
\begin{equation} 
\mu \left( \omega \right) =-{\frac{1}{\pi }}{\rm Im}\sum_{k_{1}k_{2}}\langle 
b|d^{\dagger }|k_{1}\rangle \langle k_{1}|g_{\rm eff}
\left( \omega +E_{c}\right) 
|k_{2}\rangle \langle k_{2}|d|b\rangle ,  \label{mu4} 
\end{equation} 
where $g_{\rm eff}\left( \omega \right) $ is an ``effective'' one electron 
Green's function  
\begin{equation} 
\langle k_{1}|g_{\rm eff}\left( \omega \right) |k_{2}\rangle =\langle \Phi 
_{0}|c_{k_{1}}\frac{1}{\omega -\left( H_{v}^{\prime }-E_{0}^{\prime }\right) 
+i\gamma }c_{k_{2}}^{\dagger }|\Phi _{0}\rangle .  \label{gtrans} 
\end{equation} 
The quantity $E_{0}^{\prime }$ is the ground state energy of $H_{v}^{\prime 
} $ for $N_{v}$ electrons, and $E_{c}=\varepsilon 
_{c}+E_{0}^{0}-E_{0}^{\prime }$ is the renormalized core electron energy. 
This Green's function is not of standard form since $\left| \Phi 
_{0}\right\rangle $ is an eigenfunction of $H_{v}$ and we have $%
H_{v}^{\prime }$ in the denominator. 
 
The theory developed so far is quite general, and can even account for 
losses, threshold singularity effects, and deviations from the ``final state 
rule,'' i.e., the prescription that the XAS is given by a one-electron 
expression with dipole matrix elements between the initial core and final 
state wave functions calculated in the presence of the core hole. The 
function $\mu \left( \omega \right) $ is always positive, and $g_{\rm eff}$ can 
be written in terms of a Hermitian spectral function. An approximation for 
the ``transient'' Green's function in Eq.~(\ref{gtrans}) to describe the 
edge shape was given in Ref.\ \onlinecite{Almbladh83}, pp 674. However here 
we want to describe loss processes, and thus we have to develop different 
approximations. These approximations, as discussed below, can be summed 
up in the quasi-boson representation described in Sec.\ II C. 
 
To get a qualitative feeling for the properties of the transient Green's 
function, we first discuss them in the Hartree-Fock approximation. The 
ground state of $H_{v}$, $\left| \Phi _{0}\right\rangle$  then is a Slater 
determinant, and the intermediate states in Eq.~(\ref{gtrans}) are Slater 
determinants built from orbitals which are self-consistent solutions of a 
Schr\"{o}dinger equation with the core-hole Hamiltonian $H_{v}^{\prime 
}=H_{v}+V$. We are free to choose any complete set to represent the states $%
k $ in the optical transition operator $\Delta =\Sigma _{k}\langle 
k|d|b\rangle c_{k}^{\dagger }b$. Here, we take the states that belong to $%
H_{v}$ which have the convenient property that $c_{k}^{\dagger }\left| \Phi 
_{0}\right\rangle =0$ for $k<k_{F}$. In $H_{v}^{\prime }$ we single out one 
term $h^{\prime }$ with orbitals that describe the photoelectron, and one $%
H_{v0}^{\prime }$ with orbitals for the rest of the system,  
\begin{equation} 
H_{v}^{\prime }=h^{\prime }+H_{v0}^{\prime }.  \label{Hvprime} 
\end{equation} 
These two terms both have interactions with the core hole, $V_{pc}$ and $%
V_{vc}$ respectively, which are screened versions of $V$ in Eq.~(\ref{Vc}), 
since we consider a self-consistent HF solution for the core hole 
Hamiltonian. In terms of the old Hamiltonian $H_{v}=h+H_{v0}$,  
\begin{equation} 
h^{\prime }=h+V_{pc},\;\;\;\;H_{v0}^{\prime }=H_{v0}+V_{vc}.  \label{VvcVpc} 
\end{equation} 
With the $h^{\prime }$ orbitals separated from the $H_{v0}^{\prime }$ 
orbitals, only the one-electron operator $h^{\prime }$ can couple $%
c_{k_{2}}^{\dagger }$ to $c_{k_{1}}^{{}}$ in Eq.~(\ref{gtrans}). In such a 
case we can use a product space, $c_{k}^{\dagger }\left| \Phi 
_{0}\right\rangle =\left| \Phi _{0}\right\rangle \left| k\right\rangle $, 
which then gives for the many-body XAS,  
\begin{equation} 
\mu (\omega )=-{\frac{1}{\pi }}{\rm Im}\,\langle b|d^{\dagger 
}Pg_{\rm eff}(\omega +E_{c})Pd|b\rangle ,  \label{muprime} 
\end{equation} 
where  
\begin{equation} 
g_{\rm eff}(\omega )=\left\langle \Phi _{0}\left| \frac{1}{\omega -\left( 
H_{v}^{\prime }-E_{0}^{\prime }\right) +i\gamma }\right| \Phi 
_{0}\right\rangle , 
\end{equation} 
and the projection operator onto unoccupied one-particle states of the 
initial state (without a core hole) is  
\begin{equation} 
P=\sum_{k>k_{F}}|k\rangle \langle k|.  \label{projop} 
\end{equation} 
The separation of $H_{v}^{\prime }$ into $h^{\prime }$ and $H_{v0}^{\prime }$ 
is clearly an approximation. However, it makes physical sense and has been 
used previously, e.g.\ for photoemission problems \cite{Hedin99}. In this 
expression the coupling between the photoelectron and the valence electrons 
is not present, since it is a correlation effect beyond the HF 
approximation. This defect is of less importance for a localized system, 
where HF theory is often quite useful. 
 
Let us now turn to the correlated case. A correlated $\left| \Phi 
_{0}\right\rangle $ can in principle be calculated from configuration 
interaction theory. Then $\left| \Phi _{0}\right\rangle $ is a sum of Slater 
determinants having differing numbers of electron-hole excitations (virtual 
excitations). The Slater determinants with virtual states close to the Fermi 
level have the largest coefficients in this expansion. For photoelectron 
states $k$ away from this virtual cloud it is a good approximation to use a 
product space. For definiteness we will use  
\begin{equation} 
c_{k}^{\dagger }|\Phi _{0}\rangle \approx \left\{  
\begin{array}{c} 
|\Phi _{0}\rangle |k\rangle ,\;k>k_{F} \\  
0,\;\ \ \ \ \ \ \ \ k<k_{F}, 
\end{array} 
\right.   \label{prodspace} 
\end{equation} 
being aware that this is a dangerous approximation for $k\simeq k_{F}$. As 
before we split $H_{v}^{\prime }$ into one part $h^{\prime }$ which 
describes the photoelectron, and one part $H_{v0}^{\prime }$ for other 
excitations of the valence electrons, and we introduce core hole potentials 
as in Eq.~(\ref{VvcVpc}). In addition we now also have the dynamic coupling $%
V_{pv}$ between the photoelectron and the valence electrons,  
\begin{eqnarray} 
V_{pv} &=&\sum_{k_{1}k_{2}}\sum_{l_{1}l_{2}}^{val}\left\langle 
k_{1}l_{1}\left| \left| v\right| \right| k_{2}l_{2}\right\rangle 
c_{k_{1}}^{\dagger }c_{k_{2}}^{{}}  \nonumber \\ 
&&\qquad \times \left[ {c_{l_{1}}^{\dagger }c_{l_{2}}^{{}}-\langle 
c_{l_{1}}^{\dagger }c_{l_{2}}^{{}}\rangle }\right] .  \label{Vpv} 
\end{eqnarray} 
The term $\left\langle k_{1}l_{1}\left| \left| v\right| \right| 
k_{2}l_{2}\right\rangle $ is an antisymmetrized matrix element of the 
Coulomb potential $v({\bf r})$, and the expectation value $\langle 
c_{l_{1}}^{\dagger }c_{l_{2}}^{{}}\rangle $ is subtracted, since it is 
already included in the definition of $h^{\prime }$. The state $|\Phi 
_{0}\rangle $ is an eigenfunction of $H_{v0}$ built from one-electron 
eigenfunctions without core hole potential while $h^{\prime }$ is built from 
eigenfunctions $|{k^{\prime }}\rangle $ with a core hole present,  
\begin{equation} 
h^{\prime }=\sum_{{k^{\prime }}>k_{F}}\varepsilon _{k}\,|{k^{\prime }}%
\rangle \langle {k^{\prime }}|.  \label{hstar} 
\end{equation} 
The states $|{k^{\prime }}\rangle $ are scattering states. There is thus a 
one to one correspondence between $|{k^{\prime }}\rangle $ and $|k\rangle $, 
and the energies are unchanged. Since $h^{\prime }$ only has terms bilinear 
in the photoelectron operators $c_{k}$, and since there is a linear relation 
between the states $|{k^{\prime }}\rangle $ and $|k\rangle $, $h^{\prime }$ 
will not take us outside the product space $|\Phi _{0}\rangle |k\rangle $. 
 
From Eqs.~(\ref{mu3}) and (\ref{prodspace}), we again obtain Eq.\ (\ref 
{muprime}) for the XAS, but with $g_{\rm eff}(\omega )$ in Eq.~(\ref{mu4}) 
replaced by  
\begin{equation} 
g_{\rm eff}(\omega )=\left\langle \Phi _{0}\left| \frac{1}{\omega -\left( 
H_{v0}^{\prime }-E_{0}^{\prime }\right) -h^{\prime }-V_{pv}+i\gamma }\right| 
\Phi _{0}\right\rangle .  \label{gtrans2} 
\end{equation} 
Eqs.~(\ref{Hvprime}), (\ref{VvcVpc}), (\ref{muprime}) and (\ref{gtrans2}) 
form the basis for our analysis of x-ray absorption. It is easy to show that 
Eq.~(\ref{muprime}) gives a non-negative absorption cross section $\mu 
(\omega )$, as it should. 
 
\subsection{Limiting cases} 
 
We start our analysis of the theoretical model developed above by discussing 
the two limiting cases: 1) when there is no extrinsic scattering ($V_{pv}=0$%
), and 2) when the core hole potential is neglected ($H_{v0}^{\prime}=H_{v0}$%
). 
 
\subsubsection{\it No extrinsic scattering} 
 
In this case $V_{pv}=0$, and we can put in a complete set of eigenstates $%
|\Phi _{n}^{\prime }\rangle $ to $H_{v0}^{\prime }$ with eigenvalues $%
E_{n}^{\prime }$ and obtain (taking $\gamma $ as an infinitesimal)  
\begin{equation} 
g_{\rm eff}(\omega )=\sum_{n}\frac{|\langle \Phi _{0}|\Phi _{n}^{\prime }\rangle 
|^{2}}{\omega -\omega _{n}-h^{\prime }+i\gamma },  \label{geffin} 
\end{equation} 
where $\omega _{n}=E_{n}^{\prime }-E_{0}^{\prime }$. Putting in eigenstates $%
|{k^{\prime }}\rangle $ of $h^{\prime }$, and taking the imaginary part, we 
then obtain,  
\begin{eqnarray} 
\mu (\omega ) &=&\sum_{k,n}|\langle \Phi _{0}|\Phi _{n}^{\prime }\rangle 
|^{2}|\langle {k^{\prime }}|Pd|b\rangle |^{2}\delta (\omega +E_{c}-\omega 
_{n}-\epsilon _{k})  \nonumber \\ 
&=&\int_{0}^{\omega +E_{c}-E_{F}}\!\!d\omega ^{\prime }\,A(\omega ^{\prime 
})\mu ^{(1)}(\omega +E_{c}-\omega ^{\prime }), 
\end{eqnarray} 
where the core-hole spectral function $A(\omega )$ is  
\begin{equation} 
A(\omega )=\sum_{n}|\langle \Phi _{0}|\Phi _{n}^{\prime }\rangle |^{2}\delta 
(\omega -\omega _{n}),\;  \label{Dw} 
\end{equation} 
and the one-electron XAS is  
\begin{equation} 
\mu ^{(1)}(\omega )=\sum_{{k^{\prime }}>k_{F}}|\langle {k^{\prime }}%
|Pd|b\rangle |^{2}\delta (\omega -\epsilon _{k}).  \label{mu0w} 
\end{equation} 
A similar result was derived and discussed earlier by Rehr et al \cite 
{Rehr78}. However, an important difference in our formulation is the presence 
of the projection operator $P$ in the dipole matrix element. It is 
interesting to note that $A(\omega )$ contains the core electron edge 
singularity $A(\omega )\approx \omega ^{\alpha -1}$, where $\alpha $ is the 
singularity index, and that $\mu ^{(1)}(\omega )$ is also singular at the 
Fermi level, $\mu ^{(1)}(\omega )\approx \left( \omega -E_{F}\right) ^{\beta 
}$. This latter singularity follows from the singular behavior of the 
overlap integral $\langle {k^{\prime }}|k\rangle $ \cite{Lloyd 71}. We also 
see from Eq.~(\ref{mu0w}) that the final state rule is not strictly valid, 
except well above threshold, where $P$ can be replaced by a $\delta $%
-function and the matrix element reduces to $\langle k^{\prime }|d|b\rangle $%
. We know from photoemission \cite{HMI98} that there is little extrinsic 
scattering at threshold, where this limiting case should be a good 
representation of our basic approximation given by Eqs.~(\ref{muprime}) and (%
\ref{gtrans2}). 
 
\subsubsection{\it No core hole } 
 
When the core-hole potential is neglected, $H_{v0}^{\prime }=H_{v0}$, $%
h^{\prime }=h$, and we have (taking $\gamma $ as an infinitesimal)  
\begin{equation} 
g_{\rm eff}(\omega )=\left\langle \Phi _{0}\left| \frac{1}{\omega -\left( 
H_{v0}-E_{0}\right) -h-V_{pv}+i\gamma }\right| \Phi _{0}\right\rangle . 
\label{gtrans3} 
\end{equation} 
Now $g_{\rm eff}(\omega )$ is equivalent to a standard Green's function $g\left( 
\omega \right) $. The $k_{1}k_{2}$ representation of $g(\omega )$ is  
\[ 
\left\langle k_{1}\left| g(\omega )\right| k_{2}\right\rangle = 
\] 
\begin{equation} 
\langle 0|\left\langle \Phi _{0}\left| c_{k_{1}}^{{}}\frac{1}{\omega -\left( 
H_{v0}-E_{0}\right) -h-V_{pv}+i\gamma }c_{k_{2}}^{\dagger }\right| \Phi 
_{0}\right\rangle \left| 0\right\rangle ,  \label{gk1k2} 
\end{equation} 
where $\left| \Phi _{0}\right\rangle \left| 0\right\rangle $ is an 
eigenfunction of the full Hamiltonian\ $H_{v0}+h+V_{pv}$ since $V_{pv}\left| 
0\right\rangle =0$. We can express $g$ in terms of a spectral function $%
A(\omega )$,  
\begin{eqnarray} 
\left\langle k_{1}\left| g(\omega )\right| k_{2}\right\rangle  
&=&\int_{E_{F}}^{\infty }\frac{\left\langle k_{1}\left| A(\omega ^{\prime 
})\right| k_{2}\right\rangle d\omega ^{\prime }}{\omega -\omega ^{\prime 
}+i\gamma }  \nonumber \\ 
&=&\langle k_{1}| \frac{1}{\omega -h-\Sigma (\omega )}| 
k_{2}\rangle .  \label{gkk1} 
\end{eqnarray} 
This limiting case gives a theory very similar to one-electron theory, but 
with an additional complex, energy dependent one-electron potential $\Sigma 
(\omega )$. If we approximate $\Sigma (\omega )$ by a constant $-i\Gamma $, 
which is equivalent to a Lorentzian line shape for $A(\omega )$, we recover 
the conventional XAFS result, in which extrinsic losses are represented by a 
mean free path $\lambda _{k}\approx k/\Gamma $ term, i.e., with a factor $%
\exp (-R/\lambda _{k})$ in each multiple-scattering path of length $R$. 
In general $\Sigma \left( \omega \right) $ has structure at energies away 
from the quasi-particle energy, giving rise to satellite effects. $\frac{{}}{%
{}}$ 
 
\subsection{Quasi-boson representation} 
 
We now turn to the general case, in which all three potentials $V_{pv}$, $%
V_{pc}$ and $V_{vc}$ that couple the three subsystems -- the photoelectron, 
valence electrons, and core electron -- are nonzero. To handle this we 
introduce a quasi-boson model Hamiltonian,  
\begin{eqnarray} 
H_{v0} &=&\sum_{n}\omega _{n}a_{n}^{\dagger }a_{n}^{{}},\qquad h^{\prime 
}=\sum_{k>k_{F}}\epsilon _{k}c_{k}^{\dagger }c_{k}^{{}},  \label{Hv1} \\ 
V_{vc} &=&-\sum_{n}V_{bb}^{n}\left( a_{n}^{\dagger }+a_{n}^{{}}\right) , 
\label{Vvc} \\ 
V_{pv} &=&\sum_{nk_{1}k_{2}}\left[ V_{k_{1}k_{2}}^{n}a_{n}^{\dagger 
}+(V_{k_{1}k_{2}}^{n})^{\ast }a_{n}^{{}}\right] c_{k_{1}}^{\dagger 
}c_{k_{2}}^{{}}.  \label{Vpc} 
\end{eqnarray} 
This model together with Eqs.\ (\ref{VvcVpc}), (\ref{muprime}) and (\ref 
{gtrans2}) define the set of approximations that we use in this work. The 
potential $V_{pc}$ never appears explicitly since $h\left| \Phi 
_{0}\right\rangle =0$, and thus we do not have to worry about the transform 
between the $h$ and $h^{\prime }$ states.  
 
The quasi-boson model has been discussed e.g., in Ref.\ %
\onlinecite{Bardyszewski85} and \onlinecite{Hedin99}. The essence is that 
the electron-hole type excitations are represented by bosons $a_{n}$ with 
energies $\omega _{n}$, and the electron-charge fluctuation coupling is 
represented by a term linear in the boson operators, as in Eq.~(\ref{Vpc}). 
This is analogous to the usual electron-phonon coupling. Eq.~(\ref{Vvc}) is 
a special case of Eq.~(\ref{Vpc}) with $c_{k_{1}}^{\dagger }c_{k_{2}}^{{}}$ 
replaced by $bb^{\dagger }$ and a minus sign, because the core hole 
potential is attractive. The quantities $V^{n}$ are fluctuation potentials 
corresponding to excited states $n$. The $V^{n}$ can be obtained e.g., from 
an RPA type dielectric function \cite{HMI98}. With this simple model 
Hamiltonian we can solve explicitly for the relation between the ground 
states of $H_{v0}$ and $H_{v0}^{\prime }$, i.e.,  
\[ 
|\Phi _{0}\rangle =e^{-S}|\Phi _{0}^{\prime }\rangle ,\;S=\frac{a}{2}%
-\sum_{n}\frac{V_{bb}^{n}}{\omega _{n}}\tilde{a}_{n}^{\dagger 
},\;a=\sum_{n}\left( \frac{V_{bb}^{n}}{\omega _{n}}\right) ^{2}, 
\] 
where $\tilde{a}_{n}^{\dagger }$ belongs to $H_{v0}^{\prime }=\sum_{n}\omega 
_{n}^{\dagger }\tilde{a}_{n}^{\dagger }\tilde{a}_{n}^{{}}$. Expanding to 
second order in the coupling functions $V^{n}$, we obtain
\[ 
g_{\rm eff}(\omega )= 
\] 
\[ 
\left\langle \Phi _{0}^{\prime }\left| e^{-S^{\dagger }}{\frac{1}{\omega 
-(H_{v0}^{\prime }-E_{0}^{\prime })-h^{\prime }-V_{pv}+i\gamma }}%
e^{-S}\right| \Phi _{0}^{\prime }\right\rangle  
\] 
\begin{eqnarray} 
=e^{-a} &&\left\{ {g(\omega )+\sum_{n}\left( \frac{V_{bb}^{n}}{\omega _{n}}%
\right) ^{2}g(\omega -\omega _{n})}\right.   \nonumber \\ 
&&-2\left. \sum_{n}\frac{V_{bb}^{n}}{\omega _{n}}g(\omega -\omega 
_{n})V^{n}{}g(\omega )\right\} ,  \label{geff2} 
\end{eqnarray} 
where (cf.\ Eq.~(\ref{gtrans2})), 
 
\begin{eqnarray} 
g\left( \omega \right)  &=&\left\langle \Phi _{0}^{\prime }\left| \frac{1}{%
\omega -\left( H_{v0}^{\prime }-E_{0}^{\prime }\right) -h^{\prime 
}-V_{pv}+i\gamma }\right| \Phi _{0}^{\prime }\right\rangle   \nonumber \\ 
&\equiv &\frac{1}{\omega -h^{\prime }-\Sigma \left( \omega \right) +i\gamma }%
,  \label{qbgw} 
\end{eqnarray} 
is the damped Green's function calculated in the presence of a core hole 
potential. With the above result for $g_{\rm eff}(\omega )$ we have achieved our 
goal of expressing $\mu \left( \omega \right) $ in Eq.~(\ref{muprime}) as an 
expansion in one-particle Green's functions, thus avoiding the calculation 
of correlated many body final states. In the next section we take the 
further step of making a MS expansion of the Green's functions. We note that 
the limiting expressions in Sec.~II.~B.\ all come out the same if we had chosen
to start with the quasi-boson model. 
 
\subsection{Qualitative discussion of 2nd order expression} 
 
While our basic approximation in Eqs.~(\ref{muprime}) and (\ref{gtrans2}) 
gives positive absorption $\mu (\omega )$, there is no guarantee that the 
individual terms in an expansion in powers of the coupling functions $V^{n}$ 
should be positive. The coupling strength may be gauged by the value of the 
dimensionless coefficient $a$ defined above.
For electron gas models of solids, counting 
only plasmon modes, its value is typically is $0.2-0.4$, and hence is quite 
strong. For values of $k$ close to the Fermi surface the extrinsic 
quasiparticle strength $Z$ is about $\exp (-a)\approx 0.8-0.7$. Such values for $Z$ 
are typical for most solids. The fact that effects of order $a$ are {\it not}
generally observed in XAS can be viewed as empirical evidence for 
strong cancellation effects among the various losses. However, given these 
large values of $a$, it is not surprising to encounter some non-physical 
effects in numerical calculations, such as small regions where the spectral 
function can become negative, which are due to the neglect of terms higher 
than 2nd order in the theory. 
 
Summarizing our second order expression, and changing the definition of $%
g_{\rm eff}$ in Eq.~(\ref{gtrans2}) by taking out the $e^{-a}$ factor, 
we have for the absorption spectrum,  
\begin{equation} 
\mu \left( \omega \right) =-{\frac{e^{-a}}{\pi }}{\rm Im}\,\langle 
b|d^{\dagger }Pg_{\rm eff}(\omega +E_{c})Pd|b\rangle ,  \label{mu2ndorder} 
\end{equation} 
where  
\begin{eqnarray} 
g_{\rm eff}(\omega )=g_{qp}(\omega ) &+&g_{extr}(\omega )+g_{intr}(\omega 
)+g_{inter}(\omega ),  \nonumber \\ 
&&  \nonumber \\ 
g_{qp}(\omega ) &+&g_{extr}(\omega )=g(\omega ),  \nonumber \\ 
g_{intr}(\omega ) &=&\sum_{n}\left( \frac{V_{bb}^{n}}{\omega _{n}}\right) 
^{2}g(\omega -\omega _{n}),  \nonumber \\ 
g_{inter}(\omega ) &=&-2\sum_{n}\frac{V_{bb}^{n}}{\omega _{n}}g(\omega 
-\omega _{n})V^{n}g(\omega ),  \label{geff} 
\end{eqnarray} 
account respectively for the quasiparticle term, the extrinsic and intrinsic 
loss satellites, and the interference between them. To handle the 
one-particle propagators $g(\omega )$ we assume that $\Sigma (\omega )$ is 
diagonal in a representation with eigenfunctions $|{k^{\prime }}\rangle $ of  
$h^{\prime }$ which according to calculations with the GW approximation for 
the self energy, is not too bad \cite{Hedin99}. For simplicity we now drop 
the prime on ${k^{\prime }}$ and write  
\begin{equation} 
\left\langle k\left| g(\omega )\right| k\right\rangle \equiv g(k,\omega )=%
\frac{1}{\omega -\varepsilon _{k}-\Sigma (k,\omega )+i\gamma }. 
\label{gkomega} 
\end{equation} 
With $k$ fixed $\left\langle k\left| g(\omega )\right| k\right\rangle $ as a 
function of $\omega $ has a quasiparticle peak and some more or less 
pronounced satellite structure. For $\omega $ near the quasiparticle peak we 
obtain an asymmetric lineshape,  
\begin{equation} 
g(k,\omega ) \approx \left\langle k\left| g_{qp}(\omega )\right| 
k\right\rangle 
=\frac{Z_{k}}{\omega -E_{k}+i\Gamma _{k}},  \label{gqp0} 
\end{equation} 
where $E_{k}$ is defined from $E_{k}=k^{2}/2+{\rm Re}\,\Sigma \left( 
k,E_{k}\right) ,$ $\Gamma _{k}=|{\rm Im}\,\Sigma (k,E_{k})|$ and $%
Z_{k}=[1-\partial \Sigma (k,\omega )/\partial \omega ]_{\omega =E_{k}}^{-1}$%
. We now make an on-shell approximation, defining functions of $\omega $,  
\[ 
Z(\omega )=Z_{k},\ \Delta E(\omega )=\Delta E_{k},\ \Gamma (\omega )=\Gamma 
_{k}, 
\] 
with the relation between $k$ and $\omega $ given implicitly through  
\begin{equation} 
\omega =E_{k}.  \label{kwrelation} 
\end{equation} 
Strictly speaking $\Sigma $ is defined from $h+\Sigma =t+V_{H}+\Sigma $, 
where $t$ is the kinetic energy and $V_{H}$ the Hartree potential. Often we 
would like to use e.g.\ $V_{LDA}$ rather than $V_{H}$ to generate basis 
functions. We then have to replace $\Sigma $ by $\Sigma -V_{LDA}+V_{H}$ in 
our expressions. With this on-shell approximation we have  
\begin{equation} 
\langle k|g_{qp}(\omega )|k\rangle \approx \langle k|{\frac{Z(\omega )}{%
\omega -h^{\prime }-\Delta E(\omega )+i\Gamma (\omega )+i\gamma }}|k\rangle , 
\label{gqpkk} 
\end{equation} 
and since the $|k\rangle $ are eigenfunctions of $h^{\prime }$, the operator  
$g_{qp}(\omega )$ becomes
\begin{equation} 
g_{qp}(\omega) = \frac{Z(\omega )}{\omega -h^{\prime }-\Delta E(\omega 
)+i\Gamma (\omega )+i\gamma }.  \label{gqp} 
\end{equation} 
With this form for the quasi-particle propagator $g_{qp}(\omega )$ we can 
use the MS expansion to treat the XAFS \cite{Rehr00}.
We write $h^{\prime }=t+V_{0}^{\prime }+V_{scatt}$, where $t$ is the kinetic 
energy operator, $V_{0}^{\prime }$ the potential in the central cell with 
its core hole, and $V_{scatt}$ the total scattering potential from all the 
neighboring cells (excluding inelastic losses). In the MS expansion we can 
e.g., use $h_{0}=t+V_{0}^{\prime }-i\Gamma (\omega )-i\gamma $ to obtain the 
propagator for the central absorber and then treat $V_{scatt}$ as the 
perturbation. 
 
From Eqs.~(\ref{mu2ndorder}) and (\ref{geff}), the quasi-particle 
contribution to the x-ray absorption is  
\begin{equation} 
\mu _{qp}(\omega )=-\frac{e^{-a}}{\pi }{\rm Im}\,\langle b|d^{\dagger 
}Pg_{qp}(\omega )Pd|b\rangle ,  \label{muqp1} 
\end{equation} 
where the propagator $g_{qp}(\omega )$ is given by Eq.~(\ref{gqp}). This is 
similar to the standard one-electron formula for the x-ray absorption
$\mu^{(1)}(\omega)$ with 
mean free path effects from the damping parameter $\Gamma (\omega )$, except 
for a complex amplitude factor $Z(\omega )=\exp (i\phi )|Z(\omega )|$ with a 
many-body phase shift $\phi $ and a wave-function overlap reduction by the 
factor $e^{-a}$. With $\Gamma $ and $a$ both zero and $Z=1$, the 
quasi-particle XAS $\mu _{qp}\left( \omega \right) $ becomes identical to 
the one-particle absorption $\mu ^{(1)}(\omega )$ in Eq.~(\ref{mu0w}). 
 
To evaluate the total absorption $\mu (\omega )$ including intrinsic losses 
and interference, we can, correct to terms of 2nd order in the fluctuation 
potentials $V^{n}$, replace $g(\omega )$ by $g_{qp}(\omega )$ in $%
g_{intr}(\omega )$ and $g_{inter}(\omega )$. For definiteness we define the extrinsic 
satellite function $g_{extr}(\omega )$ as the difference between the full 
propagator $g\left( \omega \right) $ and the quasi-particle propagator $%
g_{qp}\left( \omega \right) $ (see Appendix A). As noted above, the 
projection operator $P$ in Eq.~(\ref{mu2ndorder}) and Eq.\ (\ref{mu0w}) is 
necessary to account for edge singularity effects, but does not significantly
affect the fine structure. In Appendix A we also show that the many-body
expression for the x-ray absorption $\mu (\omega )$ can be expressed as a
convolution of an effective spectral function
$A_{\rm eff}(\omega ,\omega ^{\prime })$, and the quasi-particle absorption
$\mu_{qp}$ from Eq.~(\ref{muqp1}), i.e.,  
\begin{equation} 
\mu (\omega -E_{c})=\int d\omega ^{\prime }A_{\rm eff}(\omega ,\omega ^{\prime 
})\,\mu _{qp}(\omega -\omega ^{\prime }),  \label{mufinal} 
\end{equation} 
where $\omega ^{\prime }$ is the excitation energy. 
 
To evaluate our theory numerically in real systems is a heavy undertaking, 
and we will here only carry out some rough estimates which illustrate the 
theory and yield non-negligible corrections to the usual XAFS procedure \cite 
{Rehr00}. To simplify these calculations we rely on electron gas theory
within the plasmon-pole approximation to evaluate the various contributions
to the effective spectral function, and then use these results to estimate
the corresponding contributions to $\mu (\omega )$.  
Since our aim here is only to carry out a pilot study, which
is at best semi-quantitative, such an approximate model seems appropriate.
However, our approach is more general, and could be refined at the expense
of much heavier calculations.

\section{ Model Calculations} 
 
In this section we present electron gas model calculations based on the
plasmon-pole approximation (Appendix B) for the various 
contributions to the effective spectral function $A_{\rm eff}$. They can be 
represented as a sum of quasiparticle, interference, intrinsic, and 
extrinsic satellite terms defined in Appendix A, i.e.,  
\begin{equation} 
A_{\rm eff}(\omega ,\omega ^{\prime })=[1+2a(\omega )]\delta (\omega ^{\prime 
})+A^{sat}(\omega ,\omega ^{\prime }), 
\end{equation} 
where  
\begin{eqnarray} 
A^{sat}(\omega ,\omega ^{\prime }) &=&A_{extr}(\omega ,\omega ^{\prime 
})+A_{intr}(\omega ,\omega ^{\prime })  \nonumber \\ 
&&\qquad -2A_{inter}^{sat}(\omega ,\omega ^{\prime }). 
\end{eqnarray} 
Since we make comparisons to XAFS experiments for fcc Cu metal, we have set 
$r_{s}=1.80$, which corresponds to the mean interstitial electron density.
For this density, the dimensionless constant $a$ relating the 
strength of the electron-plasmon coupling to the plasmon excitation energy 
is $0.31$. Near threshold the net weight of each 
of the contributions $A_{extr},\;A_{inter}^{sat}$ and $A_{intr}$ is
equal to $a$ and their shapes are similar, so that the sum of all of
these contributions tend to cancel.  Also near threshold, 
$Z\approx \exp (-a)$ and the interference contribution to the quasi-particle 
peak $a(\omega )\approx a$. Thus the net strength of the main peak at 
threshold in our 2nd order theory is
$Z\exp (-a)(1+2a)\rightarrow 1 + O(a^{2})$. 
 
\begin{figure}
\includegraphics{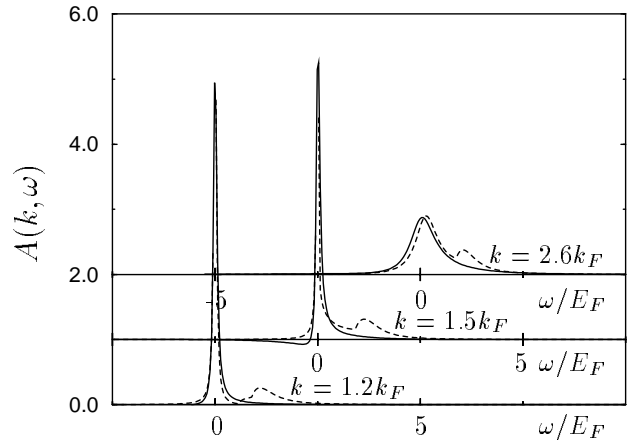}
\caption{\label{fig1}
Quasiparticle spectral function $A_{qp}^{{\rm mod}}(k,\omega +E_{k})$
in Eq.~(\protect\ref{Aqpmod}) plotted vs $\omega $ for different values
of the wave number $k$ (solid line); and
the spectral function (cf.\ Eq.~(\protect\ref{gkomega})
$(-1/\pi)\, {\rm Im}\, g$ (dashed line).  All parameters
are calculated using a plasmon pole dielectric function for an electron gas
at the mean interstitial electron density in Cu $r_{s}=1.80$.
}
\end{figure}

\begin{figure}
\includegraphics{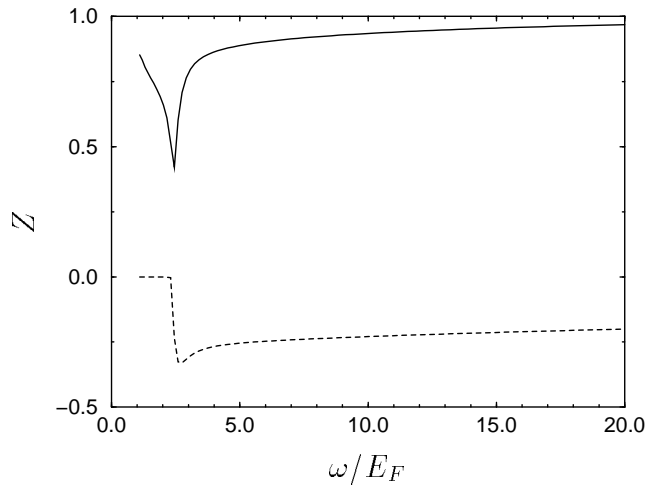}
\caption{\label{fig2}
Real (solid line) and imaginary (dashed) parts of the quasiparticle
renormalization
constant $Z(\omega )=1/(1-\partial \Sigma /\partial \omega )$, calculated at
the quasiparticle peak, for the GW plasmon pole self-energy $\Sigma$
of an electron gas at $r_{s}=1.80$. The sharp structure occurs at the
onset of plasmon excitations.
}
\end{figure}

As noted above, the asymmetric {\it quasiparticle} spectrum $A_{qp}\left( 
k,\omega \right) =(-1/\pi )\,{\rm Im}\,g_{qp}(k,\omega )$ [see Eq.~(\ref 
{gqp0})], gives rise to a net reduction in the XAFS amplitude as well as an 
additional phase shift compared to one-electron theory. These effects are 
due to the behavior of the complex renormalization constant $Z(\omega )$, 
which gives $A(\omega )$ an asymmetric Fano lineshape. In Fig.~\ref{fig1}
we plot the 
modified quasi-particle spectrum $A_{qp}^{{\rm mod}}(k,\omega)$,
where long range contribution from the imaginary part
of $Z_{k}$ is cut off in $A_{qp}^{{\rm mod}}$,
(cf.\ Appendix A and Eq.~(\ref{gqpmod})) i.e.,  
\begin{equation} 
A_{qp}^{{\rm mod}}(k,\omega +E_{k})=\frac{1}{\pi }{\frac{{\Gamma _{k}{\rm Re}%
\,Z_{k}-\omega e^{-(\omega /2\omega _{p})^{2}}{\rm Im}\,Z_{k}}}{{\omega }%
^{2}+\Gamma _{k}^{2}}}.  \label{Aqpmod} 
\end{equation} 
For comparison we also show the total extrinsic spectral function including
both quasiparticle and satellite parts.
Note that $A_{qp}^{{\rm mod}}(k,\omega )$ has nothing to do with the 
different contributions to $A_{\rm eff}$. The real and imaginary parts of the 
renormalization constant $Z(\omega )$ are plotted in Fig.~\ref{fig2}.

\begin{figure}
\includegraphics{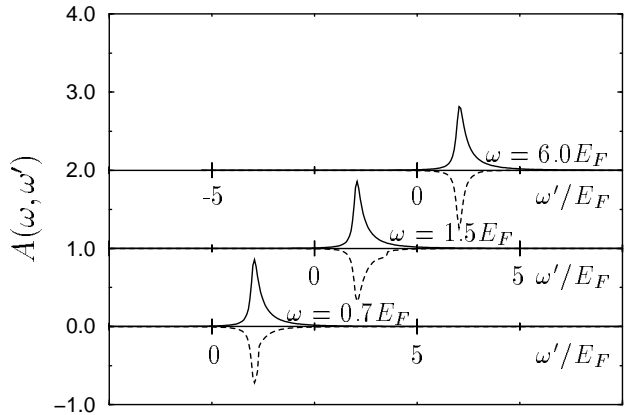}
\caption{\label{fig3}
Intrinsic (solid line) and interference satellite
(dashed) spectral functions
$A_{intr}(\omega,\omega')$ and $A_{inter}^{sat}(\omega,\omega')$,
plotted against $\omega'$ for selected values of $\omega$.
Note that the interference
contribution is negative and tends to cancel the intrinsic satellite. These
quantities are both calculated using the plasmon-pole dielectric function
for an electron gas at density $r_{s}=1.80$, and broadened by a
Lorentzian of width $0.2\omega_p$.
}
\end{figure}

The {\it intrinsic contribution} $A_{intr}(\omega,\omega^{\prime})$ to $%
A_{\rm eff}(\omega,\omega^{\prime})$ is independent of $\omega$ and gives a
well defined satellite structure peaking at an energy $\omega _{p}$ away
from the quasiparticle (Fig.~\ref{fig3}).
Although $A_{intr}$ turns on sharply, this singular 
structure is suppressed by broadening and interference terms as described 
below. The {\it interference} between extrinsic and intrinsic losses results 
in a net shift of spectral weight away from the satellite and to the 
quasiparticle peak, overall spectral weight being conserved. The rough 
cancellation of the satellite terms due to interference is clearly
illustrated by the behavior of $A_{inter}^{sat}$ in Fig.~\ref{fig3}.
Note that the 
interference satellite amplitude is maximal near threshold and slowly 
decreases with increasing energy over a range of several $\omega_p$.  

\begin{figure}
\includegraphics{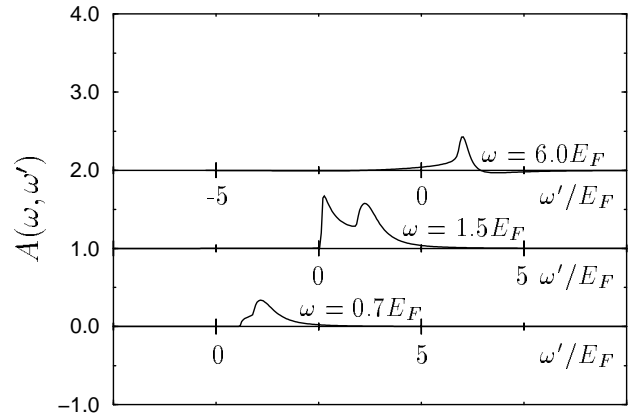}
\caption{\label{fig4}
Extrinsic satellite spectral function $A_{extr}(\omega,\omega')$
obtained as described in the Appendix for selected values of
$\omega$, vs $\omega'$.  The curve for
$\omega =0.7E_{F}$ and $\omega=1.5 E_F$ lie before the onset of
plasmon excitations. The result for $\omega=1.5 E_F$ illustrates
the anomalous structure seen in this region, and that for $\omega=6  E_F$
lies well beyond the onset. All results are based on a plasmon-pole
dielectric function with $r_{s}=1.80$.
}
\end{figure}

The behavior of the extrinsic satellite spectral function is illustrated in 
Fig.~\ref{fig4}. In the plasmon-pole, electron gas model used here, the {\it %
extrinsic satellite} spectral function sometimes exhibits a complicated 
structure. Close to the Fermi energy, the structure simplifies and consists 
of a peak at an energy about $\omega _{p}$ above the quasiparticle 
energy $\omega $ and a smooth structure that falls off gradually with 
increasing energy. For $\omega $ near the onset of plasmon losses, there is 
still a pronounced satellite peak, but there is also an additional 
``anomalous" structure near the quasiparticle peak. Indeed, it seems
ambiguous whether the structure close to the quasiparticle energy should be 
considered as part of the satellite or the main peak, as the 
structure accounts for a substantial portion of the extrinsic weight $%
[1-Z(\omega )]$ that is not included in the quasiparticle peak. This 
indicates that the anomalously low and singular behavior of $Z(\omega )$ in 
this region is partly due to the singular structure of the plasmon-pole 
approximation and largely an artifact of the {\it ad hoc} method used 
to separate the main peak and the ``satellite'' spectral function.  
Above the onset of plasmon losses, the anomalous 
structure disappears, and is replaced by a small tail which extends to the 
vicinity of the quasiparticle peak. Moreover, as the quasi-particle energy 
increases, the extrinsic satellite weight becomes progressively smaller. 

\section{Implications for XAFS} 

We can now obtain rough estimates for the effect of extrinsic and intrinsic 
losses and interference on the XAFS spectrum. In the usual MS theory \cite 
{Rehr00}, the XAFS spectrum $\chi ^{(1)}(\omega )$ is a rapidly varying 
energy dependent factor in the one-particle expression for the x-ray 
absorption,  
\begin{equation} 
\mu ^{(1)}(\omega )=\mu _{0}^{(1)}(\omega )[1+\chi ^{(1)}(\omega )], 
\end{equation} 
where $\mu _{0}^{(1)}$ is the generally smooth absorption from the central 
atom alone, in the absence of MS. The conventional (broadened) one-particle 
absorption $\mu ^{(1)}(\omega )$ is obtained from Eq.~(\ref{muprime}), with $%
g_{\rm eff}(\omega )=1/[\omega -h-i\Gamma (\omega )]$, i.e., with a damped 
one-particle propagator with mean free path effects taken into account in 
terms of $\Gamma (\omega )$. This one-particle MS theory is generally  in good 
agreement with experiment. However, there remains a residual discrepancy of
about 10\% in overall XAFS amplitudes and a systematic shift in peak positions
compared  to experiment. This shift is only partly accounted for by including
the real part of an electron gas self-energy in the one-particle propagator.

In the present theory, the many body effects of 
losses and interference can be represented as in Eq.~(\ref{mufinal}), i.e., 
as a convolution of $\mu _{qp}\left( \omega \right) $ with the effective 
spectral function $A_{\rm eff}(\omega ,\omega ^{\prime })$ including 
contributions from both primary and satellite channels, as discussed in 
Appendix A. As noted above, the difference between the behavior of $\mu 
^{(1)}(\omega )$ and $\mu _{qp}\left( \omega \right) $ is qualitatively 
minor and results primarily from a constant wavefunction overlap factor $%
\exp (-a)$, a complex renormalization factor $Z(\omega )=|Z(\omega )|\exp 
(i\phi )$, and an energy shift $\Delta E\left( \omega \right) $. The 
phasefactor $\phi $ can be absorbed by adding $\phi /2$ to the central atom 
phase factor in the XAFS formula.  
 
To extract the XAFS $\chi (\omega )$ corrected for many-body effects, we 
first use a similar MS factorization of $\mu _{qp}(\omega )=\mu 
_{qp}^{0}[1+\chi _{qp}(\omega )]$ to split off a central cell contribution $%
\mu _{qp}^{0}(\omega )$, which yields a many-body expression for the atomic 
background absorption  
\begin{eqnarray} 
\mu _{0}(\omega ) &=&\int d\omega ^{\prime }A_{\rm eff}(\omega ,\omega ^{\prime 
})\mu _{qp}^{0}(\omega -\omega ^{\prime })  \nonumber \\ 
&\approx &\mu _{qp}^{0}(\omega )\int_{-\infty}^{\omega - E_F}
 d\omega ^{\prime }A_{\rm eff}(\omega ,\omega ^{\prime }). 
\end{eqnarray} 
Here $\mu _{qp}^{0}(\omega )\sim |\langle \phi _{k}^{at}|Pd|b\rangle |^{2}$ 
is the absorption from the central atom alone, in the absence of other 
scatterers, and we have neglected the variation of $\mu _{qp}^{0}(\omega )$ 
over the dominant integration range of $\omega ^{\prime }$. This is usually 
a good approximation, since the one-particle atomic background is usually a 
smooth, monotonically decreasing function of energy. Thus one expects that 
the satellite structure in the spectral function due to many-body 
excitations will generally have minor effects on $\mu _{0}(\omega )$, which 
is consistent with experimental observations. However 
the existence of sharp atomic resonances in the one-particle absorption $\mu 
_{qp}^{0}$ may lead to exceptions. The many body XAFS function $\chi (\omega 
)=(\mu -\mu _{0})/\mu _{0}$ then becomes  
\begin{equation} 
\chi (\omega )\approx \!\int \!d\omega'\tilde{A}_{\rm eff}(\omega 
,\omega ^{\prime })\chi _{qp}(\omega -\omega'),  \label{chimb} 
\end{equation} 
where the spectral function $\tilde{A}_{\rm eff}$ is now normalized to unity,  
\begin{equation} 
\tilde{A}_{\rm eff}(\omega ,\omega')=A_{\rm eff}
(\omega ,\omega')/N(\omega ), \label{aeffnorm}
\end{equation} 
and $N(\omega )$ = $\int d\omega'\,A_{\rm eff}(\omega ,\omega')$.
With this normalization in $\chi (\omega )$, the wave function 
renormalization factor $\exp (-a)$ and the magnitude of the quasi-particle 
strength $|Z|$ cancel out, while the phase $\phi $ and the energy shift $%
\Delta E(\omega )$ both remain. 
 
\begin{figure}
\includegraphics{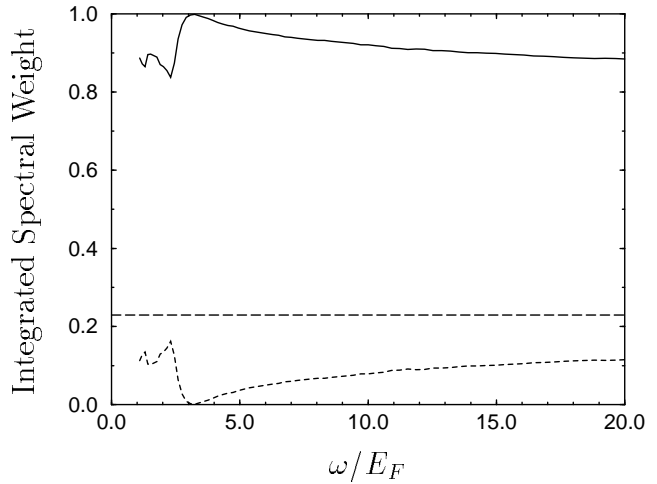}
\caption{\label{fig5}
Normalized total spectral weights of the net primary peak (solid) and
satellite terms (dashed) including interference, i.e., the integrated
quantities $\tilde{a}^{sat}(\omega )$ and $\tilde{a}_{0}(\omega )$ of
Eq.~(\protect\ref {weightsat}) and (\protect\ref{weight0}). The
anomalous structure at low energies is an artifact of the method used
to separate primary and satellite terms in the spectral function.
All results are obtained for the model of this paper with $r_{s}=1.80$.
The horizontal line represents the high energy limit of
the satellite weight (i.e., the sudden approximation).
}
\end{figure}

The importance of multi-electron excitations can be gauged by the net 
spectral weight in main peak and in the satellite structure from all losses. 
A plot of the normalized integrated satellite spectral weight,  
\begin{equation} 
\tilde{a}^{sat}(\omega )=\int d\omega^{\prime}\tilde{A}^{sat}(\omega 
,\omega^{\prime})  \label{weightsat} 
\end{equation} 
where $\tilde{A}^{sat}(\omega ,\omega^{\prime})=A^{sat}/N(\omega )$ is 
given in Fig.~\ref{fig5}.
Also plotted is the total weight of the primary peak,  
\begin{equation} 
\tilde{a}_{0}(\omega )=[1+2a(\omega )]/N(\omega ).  \label{weight0} 
\end{equation} 
Note, in particular, the slow trend of the satellite weight towards the 
sudden approximation limit [$a\exp(-a)$] with increasing energy over
a range of a few hundred eV.
In these plots we have lumped the contributions to the spectral function
that lie below the plasmon onset into the primary peak.  The 
anomalous behavior of the weights near the plasmon onset energy is due to 
the ambiguity of separating the satellite and quasiparticle contributions, 
and does not lead to singular structure in the overall absorption. 

\begin{figure}
\includegraphics{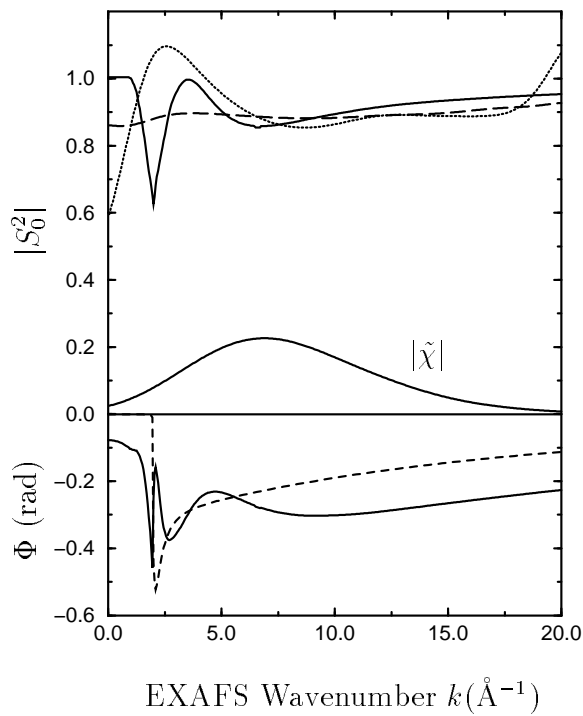}
\caption{\label{fig6}
Upper curves: magnitude $|S_0^2|$ (solid) of the XAFS many-body amplitude
reduction factor due to all inelastic losses, as calculated using a
phasor summation over the total spectral function for the model in this
work, and plotted vs the EXAFS wave number $k=\sqrt{2(\omega -E_{F})}$
for the first neighbor distance $R=2.54$ \AA\ in Cu.  Shown for
comparison are the experimental (dots) and theoretical (dashes)
amplitude reduction obtained by Fourier filtering (see text).
Note that the largest discrepancy between theory and experiment occurs
where $|\tilde{\chi}^{(1)}(k,R)|$, the Fourier filtered and back-transformed
complex XAFS amplitude, is small and experimental noise dominates.
Lower curves: phase $\Phi$ of $S_0^{2}(k,R)$ (solid), and
comparison, the many body phase shift of the asymmetric quasiparticle
peak (dashes) i.e., $\phi =\tan ^{-1}[{\rm Im}\,Z/{\rm Re}\,Z]$.
The sharp structure near $k\approx 2$ is an artifact of the sudden plasmon
onset in the plasmon pole model  used here.
}
\end{figure}

The net effect of the convolution over a normalized, positive spectral 
amplitude $\tilde{A}_{\rm eff}(\omega ,\omega ^{\prime })$ of Eq.~(\ref{chimb}) 
is clearly a decreased XAFS amplitude and a phase shifted oscillatory signal 
compared to the one-particle XAFS $\chi ^{(1)}$. In the single scattering 
approximation the oscillatory energy dependence of $\chi _{qp}(\omega )$ 
enters primarily through the complex exponential $\exp [i2k(\omega 
)R]$, where $R$ is an interatomic distance and $k(\omega )=\sqrt{2\omega }$ 
is the photoelectron wave vector. The result of the convolution can be 
written in terms of a complex amplitude factor $S_0^{2}(\omega 
,R)=|S_0^{2}(\omega ,R)|\exp (i\Phi (\omega ,R))$, which is given by an energy 
dependent ``phasor sum'' over the effective normalized spectral function,  
\begin{equation} 
S_0^{2}(\omega ,R)=\int_{0}^{\omega }
d\omega ^{\prime }\tilde{A}_{\rm eff}(\omega 
,\omega ^{\prime })e^{i2[k(\omega -\omega ^{\prime })-k(\omega )]R}. 
\end{equation} 
The qualitative behavior of $S_0^{2}(\omega ,R)$ can be understood as follows: 
At very low energies compared with the excitation energy $\omega _{p}$, the 
satellite terms strongly cancel so $A(\omega ,\omega ^{\prime })\approx 
\delta (\omega -\omega ^{\prime })$ and hence, $S_0^{2}(\omega ,R)\rightarrow 1 
$. At high energies, the sudden approximation prevails, and $A\approx 
A_{qp}+A_{intr}$, which has a strong satellite structure. However, the phase 
difference $2[k(\omega -\omega ^{\prime })-k(\omega )]$ between the primary 
channel and satellite becomes small at high energies ($\omega ^{\prime }\gg 
\omega _{p}$) and hence also $S_0^{2}(\omega ,R)\rightarrow 1$.  At intermediate 
energies, however, the value of $S_0^{2}(\omega ,R)$ has a minimum. A plot of 
the magnitude and phase of $S_0^{2}(\omega ,R)$ for our electron gas model is 
given in Fig.~\ref{fig6}., for the first neighbor
distance of Cu metal $R=2.55$ \AA. 

In order to compare these results with experiment, we isolate the first 
shell of the experimental EXAFS signal $\chi (\omega ,R)$ by Fourier filtering
over the range $1.75<R<2.80$ \AA\ in position-space $R$ (conjugate to 
$2k$), with a smooth sine window, and then back-transforming to $k$-space. 
For this comparison, we use the usual EXAFS convention for the wave number
$k=\sqrt{2(E-E_{F})}$, as measured from the threshold Fermi energy. Our 
estimate of the experimental $S_0^{2}(\omega ,R)$ is then given by the ratio 
of this back-transformed experimental first shell XAFS signal $\chi (\omega 
,R)$ to a similarly Fourier filtered and back-transformed theoretical first 
shell signal $\tilde\chi ^{(1)}(\omega ,R)$.
The latter is obtained from {\it ab 
initio} XAFS calculations using the FEFF8 code \cite{feff8}. The FEFF8 
calculations include only extrinsic losses, i.e., the mean free path loss 
calculated from a Hedin-Lundqvist plasmon-pole self-energy model. 
The results, from both theory and experiment are also plotted in
Fig.~\ref{fig6}. Due to the Fourier filtering, fine details of the
theoretical phasor sum for $S_0^2(\omega ,R)$ in Fig.~\ref{fig6}.\
are lost. Also plotted, in Fig.~\ref{fig6}.\ is the 
back-transformed XAFS amplitude $|\chi(\omega(k),R)|$ for the first nearest 
neighbor of Cu. This illustrates both where the amplitude reduction is 
important in analysis, and also where $\chi $ is small, and hence the 
$S_0^2(\omega ,R)$ extracted from experiment may have significant errors due
to experimental noise. Given the rough, electron-gas approximations used in 
our model calculations, the overall agreement with experiment for Cu metal 
is reasonably good. This result also suggests that the conventional 
procedure of approximating $S_0^2(\omega ,R)$ by a constant $S_0^2\approx 
0.9 $ is not unreasonable. The biggest discrepancies are at low energies, 
and are likely due both to experimental noise and to the approximation used 
for the mean-free path at low energies in FEFF8, which often has too
much loss.  We have also plotted the 
many-body correction to the XAFS phase, which varies by about +0.2 radians 
over the XAFS experimental range $3 < k < 20$ \AA$^{-1}$. By comparing with 
the phase of the renormalization constant $Z$ (Fig.~\ref{fig2}),
one sees that much 
of the phase shift arises from the asymmetry of the quasiparticle peak.
The sign of the phase shift is consistent with a reduction in the strength
of the self-energy due to cancellation effects.  The 
approximate linear variation of the phase with $k$ can lead to errors in 
distance determinations from XAFS measurements of about $\delta R = \Delta 
\Phi/2\Delta k \approx +0.006$ \AA\, which is comparable to systematic 
errors typically encountered in experimental XAFS analysis. 

\begin{figure}
\includegraphics{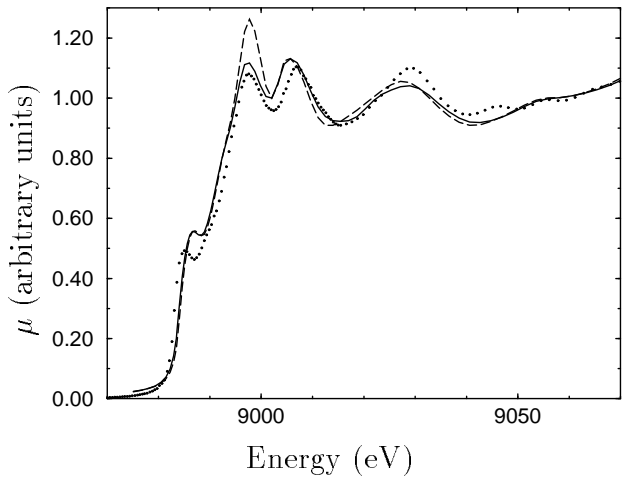}
\caption{\label{fig7}
Comparison of the full calculated many-body XAS $\mu(\omega)$  (solid)
obtained by convoluting the spectral function $A_{\rm eff}$ for
the model of this paper
with the one-particle XAS $\mu^{(1)}(\omega)$ (dashes) calculated
using the FEFF8 code, and the result from Cu experiment (dots).
}
\end{figure}

Finally we plot in Fig.~\ref{fig7} a comparison with experiment of the
full absorption result $\mu(\omega)$ obtained by convoluting the spectral
function $A_{\rm eff}$ with the quasi-particle
result $\mu_{qp}(\omega)$. These calculations were carried out
using full-multiple scattering calculations for a cluster of 300 Cu atoms
using the FEFF8 code.  To compensate for errors
in our 2nd order expression for the spectral function, $A_{\rm eff}$
was normalized to unity in this convolution; also since $\mu_{qp}$ is
real, the asymmetry in the quasiparticle peak in Eq.~(\protect\ref{Aqpmod})
was represented by a term $P\, {\rm Im}\, Z/\omega$ (where $P$ denotes the
principal part) rather than a complex phase factor $\exp{i\phi}$.  Although
it is not clear how reliable such a calculation is at the edge, given the
simplicity of our model calculations, the agreement with experiment does
markedly improve.

\section{Concluding remarks} 
 
We have developed a semi-quantitative theory for the effects of extrinsic 
and intrinsic losses and the interference between them on x-ray absorption 
spectra. The theory is based on a quasi-boson representation for the 
excitations, and perturbation theory to 2nd order in the electron-boson 
coupling. These losses lead to an asymmetric broadening of the main 
quasi-particle peak, plus a broad energy-dependent satellite in the spectral 
function due to the boson excitation. We find that the interference terms 
strongly suppress the satellite terms and enhance the main quasi-particle 
peak. These results thereby explain the general weakness of multi-electron 
excitations in XAS. We have applied the theory within the electron gas 
approximation to estimate the many-body corrections to the XAFS. By using a 
phasor summation over the spectral function, the theory yields an 
approximation to the reduction in the XAFS amplitude beyond the usual 
extrinsic, mean-free-path, together with an additional many-body phase 
shift. Pilot model calculations based on the electron gas approximation
and our formalism are in semi-quantitative agreement with XAFS experiment. 
 
Our results illustrate a striking difference with those for losses in 
photoemission. In core electron photoemission for metals there is also a 
strong extrinsic contribution to the losses for all energies, and a strong 
interference effect between extrinsic and intrinsic losses up to energies in 
the keV region \cite{HMI98}. In contrast, we found the extrinsic losses to 
have a very small influence on XAS. This at first seems to be a paradox, 
since photo absorption is usually considered to be proportional to the 
photoyield. This is indeed true in one-electron theory, and is also observed 
to be approximately true experimentally in many cases. There is, however, no 
theoretical justification for assuming this to be the case in general. If we 
examine the problem formally, the XAS is related to the dielectric 
response function (i.e., the density-density correlation function), while the
PES is given by a three current correlation function, and there are no 
simple connections between these quantities. More physically we can point at 
the fact that in PES we measure a sharp, well-defined photo electron; i.e., 
we can describe PES\ by the golden rule with well defined final states. In 
XAS the photo electrons are never measured, and photo-electron states are 
not sharp, they are decaying states with a finite lifetime. Further all 
electrons which are photo excited do not leave the solid, and some PES\ 
current which leaves the solid is due to secondary electrons which cannot 
come back and interfere at the photo excited atom. We never consider 
these secondary electrons since we do not allow our quasi-bosons to decay. 
If we could put in well-defined intermediate states with a photo electron in 
the density-density correlation function we would indeed have the same 
matrix elements 
that enter in PES, but that is not possible except perhaps approximately, 
very close to threshold when the quasi-particle life-time is long. 
 
Finally we would like to comment on similarities and differences between our 
approach and a model studied by Schrieffer \cite{Schrieffer80}. That work 
considered an Anderson-Newns like model for PES from a valence level on an 
adsorbed atom, coupled to surface plasmons. Schrieffer studied 
photoabsorption, and by cutting the polarization diagram, he identified a 
PES final state and obtained a perfect square expression for the PES 
current, including interference effects. His results agree precisely with 
what we have in our model both for XAS and when applied to PES. However the 
results for XAS cannot be expressed as a perfect square, as needed to have a 
close correspondence with PES. This is due to the differences in the signs 
of the imaginary parts, as can be seen by comparing Eq.~(7), (13) and (14a) 
in Ref.~\onlinecite{Schrieffer80}. Thus photobsorption and photoyield are 
only approximately related, even for a finite system where the 
quasi-particle aspect does not enter. 
 
\begin{acknowledgments}
We thank T. Fujikawa for useful remarks and G. Strinati 
for critical comments long ago when two of us (WB and LH) started thinking 
about this problem. This work was supported in part by DOE Grant 
DE-FG03-97ER45623/A000 (LWC and JJR). 
\end{acknowledgments}
 


\appendix
 
\section{Derivation of the Effective Spectral Function} 
 
In this Appendix we derive expressions for the different parts in the 
spectral weight function $A_{\rm eff}\equiv A_{qp}+A_{extr}+A_{intr}+A_{inter}$ 
that appears in the convolution expression for the x-ray absorption in Eq.~(%
\ref{mufinal}). Starting from Eq.~(\ref{mu2ndorder}), we have  
\begin{equation}
\mu (\omega -E_{c})=-{\frac{e^{-a}}{\pi }}\sum_{k}\left| \langle 
b|d^{\dagger }P|k\rangle \right| ^{2}{\rm Im}\,g_{\rm eff}(k,\omega ),
\label{mu2ndorder2} 
\end{equation}
provided $g_{\rm eff}$ (see Eq. \ref{geff}) is diagonal in the one-particle 
index $k$. From $GW$ calculations we know that $g$ and $g_{qp}$ are 
approximately diagonal \cite{Hedin99}, and this is hence a reasonable 
approximation for all contributions to $g_{\rm eff}$ except $g_{inter}$, which 
requires a special treatment as described later in this Appendix. We first 
introduce the real spectral weight function $A_{\rm eff}(k,\omega )$ by 
representing ${\rm Im}\,g_{\rm eff}(k,\omega )$ as  
\begin{equation} 
{\rm Im}\,g_{\rm eff}(k,\omega )=\int d\omega ^{\prime }A_{\rm eff}(k,\omega 
^{\prime })\,{\rm Im}\,g_{qp}(k,\omega -\omega ^{\prime }).  \label{aeffdef} 
\end{equation} 
Next we make the on-shell approximation of replacing $k$ in $%
A_{\rm eff}(k,\omega ^{\prime })$ by $k=k(\omega -\omega ^{\prime })$, with $%
k(\omega )$ defined in Eq. (\ref{kwrelation}). With $A_{\rm eff}\left( \omega 
,\omega ^{\prime }\right) \equiv A_{\rm eff}(k\left( \omega -\omega ^{\prime 
}\right) ,\omega ^{\prime })$ depending only on $\omega $ and $\omega 
^{\prime }$, we can perform the summation over $k$, and obtain  
\begin{equation} 
\mu (\omega -E_{c})=\int d\omega ^{\prime }A_{\rm eff}(\omega ,\omega ^{\prime 
})\,\mu _{qp}(\omega -\omega ^{\prime }),  \label{xasconv} 
\end{equation} 
where $\mu _{qp}\left( \omega \right) $ is defined in Eq.~\ref{muqp1}. Thus
the quasi-particle contribution to $A_{\rm eff}\left( \omega ,\omega ^{\prime 
}\right) $ is simply a delta-function, $A_{qp}(\omega ,\omega ^{\prime 
})=\delta (\omega ^{\prime })$. 
 
For the {\it intrinsic contribution}, it is clear from Eq.~(\ref{geff}) that  
$A_{intr}$ is simply a sum of energy-shifted $\delta $-functions,  
\begin{equation} 
A_{intr}(\omega ,\omega^{\prime})=\sum_{n}\left( \frac{V_{bb}^{n}}{\omega 
_{n}}\right) ^{2}\delta (\omega^{\prime}-\omega _{n}),  \label{aintrwwp} 
\end{equation} 
which is independent of $\omega $. In the electron gas, plasmon pole 
approximation the explicit expression for $A_{intr}$ can easily be found 
(Ref.\ \onlinecite{Almbladh83}, pp 655),  
\begin{equation} 
A_{intr}(\omega ,\omega^{\prime})=\frac{\omega _{p}^{2}\theta 
(\omega^{\prime}-\omega_{p})}{\pi (\omega^{\prime})^{3}
q(\omega') }, 
\end{equation} 
where $q(\omega')$ is a solution to $\omega_q=\omega'$.
The intrinsic spectral function contributes only at energies near $%
\omega_{p} $ above the electron quasi-particle energy, and gives a 
well-defined satellite. 
 
The {\it interference} {\it contribution} to $g_{\rm eff}$ in Eq.~(\ref{geff}) 
contains a product $g(\omega -\omega _{n})V^{n}g(\omega )$, and since $V^{n}$ 
can transfer momentum, is not diagonal in $k$ space even when $g$ is. One 
can evaluate $g_{inter}(\omega )$, e.g., by doing MS expansions of the two 
propagators, but this does not yield a result of the form in Eq.~(\ref 
{mufinal}). To force it into that form we make some further approximations. 
First we take the fluctuation potentials as plane waves, and in a plane wave 
basis we then have $\langle k+q^{\prime }|V^{q}|k\rangle =V_{0}^{q}\delta 
_{qq^{\prime }}$. Also taking the Green's functions as diagonal in a plane 
wave basis, Eqs.\ (\ref{mu2ndorder}) and \ref{geff} give  
\[ 
\mu _{inter}(\omega )=\frac{2e^{-a}}{\pi }{\rm Im}\sum_{kq}\frac{\left| 
V_{0}^{q}\right| ^{2}}{\omega _{q}}\langle b|d^{\dagger }P|k+q\rangle  
\] 
\begin{equation} 
\times \langle k+q|g\left( \omega -\omega _{q}\right) |k+q\rangle \,\langle 
k|g\left( \omega \right) |k\rangle \langle k|Pd|b\rangle .  \label{muinter0} 
\end{equation} 
Taking the quasi-particle approximation in Eq.~(\ref{gqp0}) for $g\left( 
\omega \right) $ and neglecting the $q$-dependence in the dipole matrix 
element, we have  
\begin{eqnarray} 
&&\mu _{inter}\left( \omega \right) =\frac{2e^{-a}}{\pi }{\rm Im}\sum_{kq}%
\frac{\left| V_{0}^{q}\right| ^{2}}{\omega _{q}}\left| \langle k|Pd|b\rangle 
\right| ^{2}  \nonumber \\ 
&&\qquad\qquad\times \frac{Z_{k+q}Z_{k}}{\omega _{q}+E_{k+q}-E_{k}}  \nonumber \\ 
&&\times \left[ \frac{1}{\omega -\omega _{q}-E_{k+q}+i\Gamma _{k+q}}-\frac{1%
}{\omega -E_{k}+i\Gamma _{k}}\right] ,\quad  \label{muinter1} 
\end{eqnarray} 
where we have made the approximation that $\Gamma _{k+q}\simeq \Gamma _{k}$ 
in the denominator of the prefactor before the term in brackets. We now have 
a difference between two Green's functions, and provided we treat the 
prefactor of this difference as a real number (e.g., neglecting the 
imaginary part of $Z$) we can again write $\mu _{inter}(\omega )$ as a 
convolution with $\mu _{qp}$, i.e.,  
\begin{eqnarray} 
\mu _{inter}(\omega ) &\approx &2\bigg[a(\omega )\mu _{qp}(\omega )  
\nonumber \\ 
 && - \int d\omega ^{\prime }A_{inter}^{sat}(\omega ,\omega ^{\prime })\mu 
_{qp}(\omega -\omega ^{\prime })\bigg],  \label{muinter} 
\end{eqnarray} 
with  
\begin{equation} 
a\left( \omega \right) =\sum_{q}\left. \frac{|V_{0}^{q}|^{2}}{\omega _{q}}%
\frac{\left| Z_{k+q}\right| }{\omega _{q}+E_{k+q}-E_{k}}\right| _{k=k\left( 
\omega \right) },  \label{aomega} 
\end{equation} 
\begin{eqnarray} 
A_{inter}^{sat}\left( \omega ,\omega ^{\prime }\right)  &=&\sum_{q}\left.  
\frac{|V_{0}^{q}|^{2}}{\omega _{q}}\frac{\left| Z_{k-q}\right| }{\omega 
_{q}+E_{k}-E_{k-q}}\right| _{k=k\left( \omega -\omega ^{\prime }\right) }  
\nonumber \\ 
&\times &\delta \left( \omega ^{\prime }-\omega _{q}\right) . 
\end{eqnarray} 
 
To further simplify the calculations, we assume that $Z_{k+q}\approx Z_{k}$. 
Thus we have separated the resulting spectral function into a term 
proportional to $a(\omega )$ that adds to the quasi-particle peak, and a 
term proportional to $A_{inter}^{sat}$ that contributes to the satellite 
structure. Approximating $E_{k}$ by $k^{2}/2$, we obtain  
\begin{eqnarray} 
a(\omega ) &=& \frac {\omega_{p}^{2} |Z_{k}| } {4\pi k}
\int_{0}^{\infty } d\omega^{\prime} \frac{1}{\omega^{\prime}(\omega 
_{p}+\omega^{\prime})^{2}}  \nonumber \\ 
&\times &\log \left[ \frac{\omega _{p}+2\omega^{\prime}+kq}{\omega 
_{p}+2\omega^{\prime}-kq}\right] , 
\end{eqnarray} 
where  $q=\sqrt{2\omega^{\prime}}$ and $k=\sqrt{2\omega }$.  Similarly
\begin{eqnarray} 
A_{inter}^{sat}\left( \omega ,\omega^{\prime}\right) &=&\frac{%
|Z_{k^{\prime}}|\, \omega _{p}^{2}}{4\pi k^{\prime}(\omega^{\prime}-\omega _{p})%
{\omega^{\prime}}^{2}}  \nonumber \\ 
&\times &\log \left[ \frac{\omega _{p}+k^{\prime}(\omega )q}{\omega 
_{p}-k^{\prime}(\omega )q}\right] \theta (\omega^{\prime}-\omega _{p}),\quad 
\end{eqnarray} 
where  $q=\sqrt{2(\omega^{\prime}-\omega _{p})}$ and
$k^{\prime}=\sqrt{2(\omega -\omega^{\prime})}$.  

For the {\it extrinsic contribution}, we have from Eq.~(\ref{aeffdef}),  
\begin{equation}
{\rm Im}\left[ \frac{g_{extr}\left( k,\omega \right) }{Z_{k}}-\int \!{\frac{%
A_{extr}(k,\omega ^{\prime })d\omega ^{\prime }}{\omega -\omega ^{\prime 
}-E_{k}+i\Gamma _{k}}}\right] =0. 
\end{equation}
The integral gives an analytic function of $\omega $ in the half-plane above 
the line $\omega =E_{k}-i\Gamma _{k}$. If we also assume that $%
g_{extr}\left( k,\omega \right) =g\left( k,\omega \right) -g_{qp}\left( 
k,\omega \right) $ is analytic in this region, we can make a shift in the 
origin for $\omega $ by $E_{k}-i\Gamma _{k}+i\delta $ ($\delta \rightarrow 
0^{+}$) without crossing the analyticity line. This gives  
\begin{eqnarray} 
&&{\rm Im}\left[ \frac{g_{extr}\left( k,\omega +E_{k}-i\Gamma _{k}+i\delta 
\right) }{Z_{k}}\right.   \nonumber \\ 
&&\qquad -\left. \int A_{extr}\left( k,\omega ^{\prime }\right) \frac{%
d\omega ^{\prime }}{\omega -\omega ^{\prime }+i\delta }\right] =0, 
\end{eqnarray} 
from which we obtain  
\begin{equation} 
A_{extr}\left( k,\omega \right) =-\frac{1}{\pi }{\rm Im}\,\left[ \frac{%
g_{extr}(k,\omega +E_{k}-i\Gamma _{k}+i\delta )}{Z_{k}}\right] . 
\label{aextr} 
\end{equation} 
The analyticity of $g_{extr}\left( k,\omega \right) $ follows if it can be 
described by a sum of discrete poles or by a contour integration below $%
\omega =$ $E_{k}-i\Gamma _{k}$. This is a reasonable approximation since the 
satellite peak is broader than the quasi-particle one. If for $g_{qp}\left( 
k,\omega \right) $ we take the expression valid for $\omega $ close to the 
quasi-particle energy  
\begin{equation} 
g_{qp}\left( k,\omega \right) =\frac{Z_{k}}{\omega -E_{k}+i\Gamma _{k}} 
\label{gqpeq} 
\end{equation} 
and use it for all $\omega $, we have  
\begin{equation} 
g_{extr}\left( k,\omega \right) =\frac{1}{\omega -\varepsilon _{k}-\Sigma 
\left( k,\omega \right) }-\frac{Z_{k}}{\omega -E_{k}+i\Gamma _{k}}. 
\label{gextr} 
\end{equation} 
Since $Z_{k}$ has a fairly large imaginary part, this leads to a substantial 
Fano type asymmetry and substantial (cancelling) negative contributions in 
the spectral functions for $g_{extr}$ and $g_{qp}$. The expression to use 
for $g_{qp}$ is not well-defined except very close to the quasi-particle 
energy. To avoid a negative spectral density for the extrinsic satellite we 
introduce a Gaussian cutoff to the imaginary part of $Z_{k}$ in ${\rm Im}%
\,g_{qp}$, i.e.,  
\begin{eqnarray}
&{\rm Im}&\,g_{qp}(k,\omega +E_{k}-i\Gamma _{k}+i\delta ) \nonumber \\
&&=\frac{-\delta {\rm Re}\,Z_{k}+\omega e^{-(\omega /2\omega _{p})^{2}}
{\rm Im}\,Z_{k}}{\omega ^{2}+\delta^{2}}. 
\label{gqpmod} 
\end{eqnarray} 
If we further replace $Z_{k}$ by $\left| Z_{k}\right| $ in Eq.~(\ref{aextr}%
), we obtain for $\omega >>\delta $  
\begin{eqnarray} 
A_{extr}(k,\omega ) &=&-\frac{1}{\pi |Z_{k}|}\bigg\{\left[ \Gamma _{k}+{\rm %
Im}\,\Sigma (k,\omega +E_{k})\right]   \nonumber \\ 
&&\times 1\bigg/\bigg[\left[ \omega +\Delta E_{k}-{\rm Re}\,\Sigma (k,\omega 
+E_{k})\right] ^{2}  \nonumber \\ 
&&\qquad +\left[ \Gamma _{k}+{\rm Im}\,\Sigma (k,\omega +E_{k})\right] ^{2}%
\bigg]  \nonumber \\ 
&&-{\frac{{\rm Im}\,Z_{k}}{\omega }}e^{-(\omega /2\omega _{p})^{2}}\bigg\}, 
\label{Aextr2} 
\end{eqnarray} 
where $\Delta E_{k}={\rm Re}\,\Sigma (k,E_{k})$. We have neglected the $%
i\Gamma _{k}$ in the argument for $\Sigma $, and we have only considered $%
\omega >>\delta $ since $g_{extr}\left( k,\omega \right) $ is small around $%
\omega =0$ when the true complex quasi-particle energy is used. This result 
can be compared with the first order result from Eq.~(\ref{aeffdef}), i.e., 
the result obtained by taking $\Gamma _{k}\rightarrow 0$. With $\Gamma _{k}=0 
$ and neglecting the phase in $Z_{k}$ we have ${\rm Im}\,g_{qp}\left( 
k,\omega \right) =-\pi \left| Z_{k}\right| \delta (\omega -E_{k})$, and thus 
from Eq.\ (\ref{aeffdef}) $A_{extr}\left( k,\omega -E_{k}\right) =-1/\left( 
\pi \left| Z_{k}\right| \right) {\rm Im}\,g_{extr}(k,\omega )$. This is the 
same result as in Eq.\ (\ref{Aextr2}) with $\Gamma _{k}=0$. 
Since the GW approximation puts the extrinsic satellite at a slightly
different position than the interference and intrinsic satellite terms,
(which leads to numerical problems such as small regions where the spectral
function is negative) we shifted $A_{extr}$ to make the peaks coincide.

In summary we have found that  
\begin{equation} 
\mu \left( \omega -E_{c}\right) =\int d\omega ^{\prime }A_{\rm eff}(\omega 
,\omega ^{\prime })\mu _{qp}(\omega -\omega ^{\prime }),  \label{muconv} 
\end{equation} 
where  
\begin{eqnarray} 
A_{\rm eff}(\omega ,\omega^{\prime}) &=&\left[ 1+2a(\omega )\right] \delta 
(\omega^{\prime}) +A_{extr}(\omega,\omega^{\prime})  \nonumber \\ 
&+&A_{intr}(\omega ,\omega^{\prime}) -2A_{inter}^{sat}(\omega 
,\omega^{\prime}).  \label{aeffomomp} 
\end{eqnarray} 
 
\section{Plasmon-pole, Electron Gas Model} 
 
In this Appendix, we briefly outline some properties of the electron gas 
model used in our calculations. In order to estimate the many-body effects 
on XAFS spectra, we need to make some simplifying approximations. By choosing
to work with an electron gas model, many of the formulae in our 
model can be found analytically, greatly simplifying the calculations. We 
further choose to use a plasmon pole dielectric function.  Although the
model exhibits some non-physical singular structure and no loss at low
energies, it never the less gives mean free paths and self-energy shifts
in reasonable agreement with experiment. Under this
approximation, the fluctuating potentials $V^{n}$ are plane waves  
\[ 
V^{n}({\bf r})=V^{{\bf q}}({\bf r})=V_{0}^{{\bf q}}e^{i{\bf qr}}. 
\] 
In the case of coupling to the core hole (at ${\bf r}={\bf 0}$) this yields  
$V_{bb}^{q}=V_{0}^{\bf q}$.
According to the plasmon pole model\cite{Lundqvist67}  
\begin{equation} 
V_{0}^{\bf q}=\left( \frac{2\pi e^{2}\omega _{p}^{2}}{q^{2}\omega _{q}\Omega }%
\right) ^{1/2}, 
\end{equation} 
where $\Omega $ is the system volume.
For the case when the plasmon dispersion has the form
$\omega^2_{q}=\omega^2_{p}+\alpha q^{2}+q^4/4$, the imaginary part of $\Sigma 
_{GW}(k,\omega )$ can be obtained analytically, i.e.,  
\begin{equation} 
{\rm Im}\,\Sigma (k,\omega )=-\frac{\omega _{p}}{4k}\log \left[ \frac{%
q_{2}^{2}}{\omega _{p}+q_{2}^{2}}\frac{\omega _{p}+q_{1}^{2}}{q_{1}^{2}}%
\right] \theta (\omega -\omega _{th}), 
\end{equation} 
where $\omega _{th}$ is the threshold for plasmon excitation,  
and $q_1$ and $q_2$ are limiting values of the inequalities
$\omega_q+(q-k)^2/2-\omega < 0$ and
$\omega_q+(q+k)^2/2-\omega > 0$, and hence are
solutions to a cubic equation,
\begin{eqnarray}
&& kq^3+\left(\omega+\alpha-\frac{3}{2}k^2\right) q^2
       +(k^3-2\omega k) q  \nonumber \\
&& + \left(\omega_p^2-\omega^2+\omega k^2-\frac{k^4}{4}\right)=0,
\end{eqnarray}
with the constraints, min$(q_1,q_2)$=0, and
max$(q_1,q_2)$=0 satisfies
$\omega_q=\omega+k^2/2-k_F^2/2$.
The real part is then obtained by a Kramers-Kronig transformation,  
\begin{equation} 
{\rm Re}\,\Sigma (k,\omega )=V_{ex}\left( k\right) +\frac{P}{\pi }\int \frac{%
{\rm Im}\,\Sigma (k,\omega ^{\prime })d\omega ^{\prime }}{\omega -\omega 
^{\prime }},  \label{kktran} 
\end{equation} 
where  
\begin{equation} 
V_{ex}\left( k\right) =-{\frac{2}{\pi }}\left[ {\frac{k_{F}}{2}}+{\frac{%
k_{F}^{2}-k^{2}}{4k}}\ln \left| {\frac{k+k_{F}}{k-k_{F}}}\right| \,\right] , 
\label{vexch} 
\end{equation} 
is the energy-independent part in $\Sigma $, i.e., the Hartree-Fock exchange 
energy.  In the calculations presented in this work we have chosen
the plasmon dispersion as in Ref.~(\cite{Lundqvist67}), i.e., 
$\alpha=2/3$, which is the same as that used in the extrinsic loss
calculations in the FEFF8 code. 
 
\end{document}